\documentstyle[astrobib,aaspp4,epsfig]{article}
\def\lap{\lower.5ex\hbox{$\; \buildrel < \over \sim \;$}}
\def\gap{\lower.5ex\hbox{$\; \buildrel > \over \sim \;$}}
\begin{document}
\title{WFPC2 Observations of Massive and Compact Young Star Clusters in M31\altaffilmark{1}}
\author{ Benjamin F. Williams}
\affil{University of Washington}
\affil{Astronomy Dept. Box 351580, Seattle, WA  98195-1580}
\affil{ben@astro.washington.edu}
\author{Paul W. Hodge}
\affil{University of Washington}
\affil{Astronomy Dept. Box 351580, Seattle, WA  98195-1580}
\affil{hodge@astro.washington.edu}

\altaffiltext{1}{Based on observations with the NASA/ESA Hubble Space
Telescope obtained at the Space Telescope Science Institute, which is
operated by the Association of Universities for Research in Astronomy,
Inc., under NASA contract NAS5-26555.}

\begin{abstract} 

We present color magnitude diagrams of four blue massive and compact
star clusters in M31: G38, G44, G94, and G293.  The diagrams of the
four clusters reveal a well-populated upper main sequence and various
numbers of supergiants.  The U-B and B-V colors of the upper main
sequence stars are used to determine reddening estimates of the
different lines of sight in the M31 disk. Reddening values range from
$E_{B-V} = 0.20\pm0.10$ to $0.31\pm0.11$. We statistically remove
field stars on the basis of completeness, magnitude and color.
Isochrone fits to the field-subtracted, reddening-corrected diagrams
yield age estimates ranging from $63\pm15$ Myr to $160\pm60$ Myr.
Implications for the recent evolution of the disk near NGC 206 are
discussed.

\end{abstract}

\section{Introduction}

Studies of the stellar content of extra-galactic globular clusters
(hereafter GCs) started in the Magellanic Clouds
(e.g. \citeNP{monograph}).  These observations led to the recognition
of several blue GCs whose ages were found to range from $\sim$10$^8$ -
10$^9$ years (\citeNP{arp1958}; \citeNP{arp1959}; \citeNP{hodge1961};
\citeNP{hodge1984}; \citeNP{brocato1989}; and many others).  The
Galaxy appears not to have such young massive and compact clusters,
though other more distant spirals do \cite{larsen}.  If they are young
GCs, observations may uncover valuable clues about the formation and
evolution of GCs.  Regardless of their classification, these clusters
are certainly useful probes of the reddening in their portion of the
galaxy because they contain upper main sequence stars at a common
distance, which have nearly the same color in B-V independent of
metallicity.

Using photoelectric photometry, \citeN{vandenbergh} observed that some
of the objects in M31 classified as GCs were very blue, including the
four clusters discussed here: G38, G44, G94, and G293 (designations
are from \citeNP{sargent}).  Integrated U, B, and V photometry of
hundreds of clusters was done with photographic plates by
\citeN{battistini}, confirming the blue colors of these four and
several others.  With the Ultraviolet Imaging Telescope aboard the
Astro-1 mission, \citeN{bohlin} detected 43 M31 GCs in the
ultraviolet, including G94, and its UV flux was consistent with its
being a young cluster.  \citeN{huchra1991} completed spectroscopic
metallicity measurements for 150 GCs in M31, including G38.  Its
metallicity measurement has high uncertainty and may be significantly
underestimated because it was determined without the knowledge of its
young age.

The first M31 globular cluster resolved into stars was studied from
the ground by \citeN{heasley1988}.  With sub-arcsecond seeing
observations from Mauna Kea, they resolved the red giant branch of
G1. More recently, the populations of star clusters in M31 have been
resolved with from space (e.g. \citeNP{ajhar1996}, \citeNP{rich1996},
\citeNP{pecci1996}, \citeNP{holland1997}, \citeNP{jablonka2000}).
These studies have used images from the Wide Field and Planetary
Camera (WFPC-2) aboard the Hubble Space Telescope (HST) to construct
color-magnitude diagrams (CMDs) of M31 star clusters in order to
determine their metallicities, reddening values, and ages; in some
cases, comparisons were made between the cluster and field populations
(e.g. \citeNP{jablonka2000}).  We are using the same instrument to
make CMDs of four of the massive and compact young clusters: G38, G44,
G94 and G293.

We begin in section 2 with a description of our observations and
reduction techniques, followed in section 3 by the details of our
field subtraction technique.  In section 4 we discuss the
determination of the reddening values and luminosity determinations
for each cluster, and section 5 describes the age determinations.
Finally, in section 6 we provide a summary of our results.

\section{Observations and Data Reduction}

All observations for this project were obtained through the F336w,
F439w, and F555w filters with the WFPC-2 imager aboard HST on October
15,16,30, and 31, 1999.  Each field had 3600 seconds of exposure in
F336w, 1600 seconds in F439w, and 1200 seconds in F555w. The short
exposure times were ideal for controlling field contamination by disk
stars of M31, minimizing the severe crowding expected in these dense
clusters, and minimizing telescope time needed.

The images were well aligned so that no shifting was necessary before
combining.  They were combined using the IRAF task gcombine, with the
cosmic ray rejection algorhythm crreject, and the photometry was
extracted using the automated photometry programs DAOPHOT-II and
ALLSTAR \cite{stetson}.  DAOPHOT-II finds the stars in the image and
generates a point spread function (PSF) from the stars which are both
isolated enough to minimize neighbors contaminating the PSF
measurement and bright enough to provide high signal to noise for the
PSF measurement.  Allstar groups the stars together and carries out
the photometry in such a way as to not include counts from neighboring
stars when fitting the PSF.  With these programs, it is possible to
obtain fairly accurate photometry for crowded fields.  PSF magnitudes
were checked against aperture photometry of the most isolated stars
with the highest signal to noise in order to determine if there was an
offset between the PSF photometry and the more standard aperture
photometry; results are shown in Figure 1.  Within the errors of our
photometry there was no significant aperture correction.  Due to the
shallowness of our images, and the fluctuation of the background level
in the structured disk of M31, the accuracy of our photometry was not
high. As shown in Figure 2, the photometric accuracy was worse for the
bluer observations; the average errors were 0.16 mag in U, 0.13 mag in
B, and 0.10 mag in V.  From the data shown in Figure 1, we calculate
internal uncertainties of 0.06 mag for the F336W data, 0.10 mag for
the F439W data, and 0.02 mag for the F555W data for the best sampled,
most isolated cases (rms).

We obtained standard U, B, and V magnitudes from our photometry using the
methods, zero points, and transformation coefficients given in
\citeN{holtzman}.  We first determined instrumental magnitudes for the F336w, F439w, and
F555w filter as 
$$ 
m_{filter} = -2.5 \times log(ADU/t) + ZP_{filter}\\ \ \ \ \ \ (1) 
$$ 
where ADU is the number of counts, t is the exposure time, and
$ZP_{filter}$ is the zero point of the WFPC-2 chip for
the bandpass.  Since the transformation to U, B, and V is a function
of color, we used the F336w - F439w as a first approximation of the
U-B color and the F439w - F555w color as a first approximation of the
B-V color and iteratively solved the transformation equations.

\section{Field Contamination}

To make sure that we sampled the whole cluster, we made CMDs
containing all the stars within 11.5" of the cluster center (43 pc).
We chose this radius because it was about twice the radius where the
difference between the star density and the average star density of
the field became comparable to the fluctuations of the field star
density.  These CMDs are shown for each cluster in Figure 3.  They
show a conspicuous main sequence with a few scattered red stars.  In
order to remove field stars from our cluster CMD, we determined the
completeness of our data as a function of color and magnitude, inside
and outside the cluster.  With quantitative knowledge of the
completeness, we could calculate how many field stars our photometry
measured inside the cluster.

\subsection{Completeness Tests}

In order to do accurate field subtraction in the cluster region, we
determined how well our photometry routine detected stars in different
regions of the chip for each cluster.  We expected the completeness to
be worse at fainter magnitudes and closer to the cluster center, but
it was necessary to quantify this effect in order to determine the
correct number of field stars at each color and magnitude that were
detected inside the cluster.

We added a unique list of 500 artificial stars to each of 20 copies of
each cluster image, making a total of 10000 artificial test stars for
each cluster.  Adding only 500 stars to each image copy insured that
the crowding of the artificial stars themselves would not effect the
completeness results.  We put each image copy through the same
photometry routine that we used for our original data in order to
learn the completeness and accuracy with which the artificial stars
were recovered.  We ran more tests closer to the center of the
cluster, and more at fainter magnitudes to improve our sampling where
we thought the completeness would be the worst.

Examples of the results of our artificial star tests are shown in
Figure 4.  Dotted contours show where the completeness was less than
ten percent (0.5 and 5 percent).  The first solid contour marks ten
percent completeness, and successive contours each mark five percent
increases in completeness, so that the thick contours mark 25, 50, and
75 percent completeness.  Above 80 percent is not as systematic
because of the small number of stars at bright magnitudes far away
from the cluster.  We found that in general our photometry was greater
than 10\% complete down to $m_B \approx 25$, but completeness
decreased substantially inside of one arcsecond from the cluster
center, as expected.  The effect was quite different for the four
clusters supporting the impression that the clusters are not equally
compact; G38 and G44 are much more crowded in their
central arcsecond than G94 or G293.

\subsection{Field Subtraction}

We used the color information of our completeness tests as well as
position and brightness information to determine the field
contamination in our cluster CMDs.  We accomplished this
multi-dimensional analysis by making different CMDs of the artificial
stars for 3 different annuli around the cluster center.  We divided
each of our artificial CMDs for the three annuli into bins of color
and magnitude.  By examining the artificial stars in each bin, we
calculated the completeness for each magnitude and color in each
annulus.

Using the CMDs of the field stars for each cluster (Figure 5), we
measured the number of field stars in each color and magnitude bin.
After correcting these measurements for the completeness inside the
cluster, we removed the proper number of stars from each color
magnitude bin in the cluster CMD.  The final field subtracted and
reddening corrected CMDs for all four clusters are given in Figure 6.
(The reddening correction will be discussed in the next section.)
Notice that the fainter red stars, along with a reasonable fraction of
the main sequence stars, have disappeared.  Also interesting is the
varying number of supergiants from cluster to cluster.  For example
G38 and G94 appear to have significantly more supergiants than G44 or
G293.

\section{Determination of Reddening and Integrated Cluster Parameters}

In order to determine the reddening, we assumed that the reddening law
of M31 is the same as the reddening law in our Galaxy.  This
assumption has been true for most observations of reddening in M31
(e.g. \citeNP{walterbos}; \citeNP{nedialkov}; \citeNP{barmby2000}).
We therefore assumed that $E_{U-B} = 0.64 E_{B-V}$ \cite{textbook},
and we used the colors from the model main sequence of
\citeN{bertelli} in order to determine the reddening of our cluster
stars.

One advantage of looking at blue clusters is that they often contain
upper main sequence stars, which have B-V colors that are nearly
independent of absolute magnitude.  These stars are easily corrected
to their intrinsic colors in the B-V, U-B plane.  All of our clusters
contained a significant number of these stars.  We used all the stars
blueward of B-V = 0.5, and brighter than $m_V = 24.5$ in order to
determine our reddening correction.  With the random errors in our
photometry, however, they did not make a typical linear sequence in
the B-V, U-B plane.  Instead, they were spread in a Gaussian-like way
to the red of the main sequence line.

We determined the average and standard errors of this distribution of
stars, under the assumption that all of their true B-V colors were
nearly equal.  Using these values, we assumed an error ellipse whose
axes corresponded to the standard deviation of the B-V and U-B colors.
We then shifted the error ellipse along the reddening line until ten
percent of the area crossed the theoretical main-sequence color line,
and again until 90 percent of the area crossed the line.  We took
these limits to be the lower and upper limits of the reddening
correction at the 90 percent confidence level, and we took the shift
which centered the distribution on the model main sequence to be the
most likely value predicted by our data.  This most likely shift for
each cluster is shown in Figure 7.  Solid lines show the model colors
for stars on the main sequence (top curve), and supergiants (bottom
curve).  The dotted line shows the reddening line.  Closed boxes are
average colors of the pre-corrected upper main sequence stars, and
open boxes are the same distributions after applying our most likely
reddening correction value.  Three of our clusters had reddening
values close to the mean value of globulars in M31 determined by
\citeN{barmby2000} from a sample of 221 clusters to be $E_{B-V} =
0.22$.  The larger value for G38 is reasonable since the lower number
of field stars near that cluster may be due to a higher extinction
level.

We obtained the half-light radii and the integrated fluxes of the
clusters by measuring the average flux per pixel in the field,
{\it{including}} the field stars.  We then subtracted off this
background and measured the total flux coming from the cluster for
different aperture sizes.  At the same time, we measured the surface
brightness at each radius, in order to determine the radius at which
the difference between the surface brightness and the average surface
brightness of the field was the same as the fluctuations in the
surface brightness of the field.  We took our measurement of the total
flux inside an aperture of this radius.  We then measured the
half-light radius, the aperture size where the flux was 1/2 this total
value.

In Table 1 we give:

Column 1: The name of each cluster from \citeN{sargent}.

Column 2: Our values for the projected galactocentric distances in kpc.

Column 3: Our measured half-light radii, typical errors were $\pm$0.3
pc.

Columns 4, 5, 6: Integrated U, B, and V magnitudes from
\citeN{battistini}.  Their typical errors were $\pm$0.15.

Column 7: Our values for the reddening ($E_{B-V}$) determined using U, B, and V
stellar photometry.

Columns 8, 9, 10: The reddening corrected U, B, and V integrated
magnitudes from our data. Typical errors were $\pm$0.10.

Column 11: The absolute V magnitude assuming a distance modulus of
24.43 \cite{freedman}.  For comparison, a typical $M_V$ for a globular
cluster in our Galaxy is -7.1.

\section{Ages}

We determined ages by two different methods.  First, we used the
automated software package MATCH \cite{dolphin}.  This software uses
the artificial star results, uncertainties from the artificial star
results, the photometry of the cluster stars, and the distance to the
cluster to create color magnitude matrices (Hess diagrams).  It then
compares these Hess diagrams to scaled theoretical Hess diagrams,
based on Bertelli isochrones \cite{bertelli}, producing a modified
$\chi^2$ value for a given range of age, metallicity, and reddening
values.  Then it compares the quality of all of the fits to determine
the most likely age, metallicity, and reddening values and an
associated error for each.  The program only works with two colors, B
and V, so that the reddening values it derives may not be as reliable
as those we determined using three colors.

As an independent age determination, using the stellar evolution
models of \citeN{bertelli}, we created isochrones for different ages
and metallicities.  We then looked at the different isochrones
overplotted on our CMDs in order to determine a reasonable upper and
lower limit on the age.  The method is shown in Figure 8, where dotted
lines mark the upper and lower limits.  These were determined by
finding the age where the blue loops fell below and above the location
of the supergiants.  The solid isochrones mark our choices for the
best fits for the turnoff age.  While the turnoff age was unaffected
by the chemical composition at these young ages, the distribution of
supergiants was often more consistent with one metallicity than
another.  While we make no claims here to have robust measurements of
the metallicities for these clusters, we label the metallicities here
in order to be complete in our descriptions of the best fitting
isochrones.

The results for both techniques are shown in Table 2, which gives the
ages, metallicities and errors determined by eye using isochrone
fitting, and the ages, metallicities and errors determined by MATCH,
using statistical comparisons of Hess diagrams.  Clearly, the two
methods derived consistent ages and metallicities, but the quoted
errors from the isochrone fitting are smaller.  This is likely due to
the fact that we estimated the isochrone errors assuming that our
reddening value is correct, whereas the uncertainties in the reddening
measurements were folded into the errors that MATCH derived.

\citeN{magnier1997} studied the young stellar population and Cepheid
variable population of this region of the M31 disk in detail.  They
found that the average stellar ages in the region south of NGC 206
($\sim$90 Myr) are slightly older than NCG 206 ($\sim$30 Myr) but
younger than the population to the north of NGC 206, suggesting that
the density wave interaction that is coincident with the location of
NGC 206 has been propagating through the galaxy at $\sim$32 km/s.  G38
and G44 are located near the center of the overdensity of Cepheids
that they find to the south of NGC 206.  Our ages for these massive
clusters show that they were likely formed at the same time as the
Cepheids in the surrounding region.  Assuming a distance modulus to
M31 of 24.43 \cite{freedman}, these clusters are located at projected
distances $\sim$3500 pc and $\sim$2600 pc south-southwest of NGC 206,
respectively.  Therefore, these ages for G38 and G44 strengthen the
argument that the spiral arm interaction possibly responsible for the
creation of NCG 206 has been propagating through M31 from south to
north at a projected speed of $\sim$32 km/s with respect to the stars
in the disk.  In fact, these ages are consistent with the possibility
that these clusters were formed by the same mechanism that has caused
the formation of NGC 206.

\section{Conclusions}

We have constructed CMDs for four massive and compact star clusters in
the disk of M31.  The CMDs reveal an upper main sequence stellar
population showing that these are indeed young clusters.  The
reddening values determined using the color of the upper main sequence
stars are comparable to typical reddening values determined for
globular clusters in the M31 halo, suggesting that three of these
clusters may be closer to the near side of the disk.  The fourth case
shows higher reddening, suggesting that it is located deeper in the
disk.  None of the reddening values are extremely high, which suggests
that these clusters have formed in regions of relatively low dust
content.

Age determinations from statistical matching of CMD properties as well
as those from isochrone fitting give consistent ages for all four
clusters.  These ages range from $63\pm15$ Myr to $160\pm60$ Myr.  The
ages of G38 and G44 show that they formed at the same time as the
well-studied surrounding Cepheid population, strengthening the
conclusions of {\citeN{magnier1997}.  The clusters' luminosities and
half light radii imply that they are more massive and compact than any
young clusters in our Galaxy; however, their luminosities are quite
similar to young clusters found in many other galaxies \cite{larsen}.
Apparently, these massive and compact young clusters are common to
most galaxies, including the Magellanic Clouds and M31, and it is very
curious that our Galaxy contains no similar clusters.

\section{Acknowledgments}

We thank Ted Wyder for his help with the analysis methods; Andrew
Dolphin for helping with the MATCH package; Andrew Becker for helping
with the reduction and analysis programming; and Frank van den Bosch
for helping to create some of the plots.  Support for this work was
provided by NASA through grant number GO-06459.01-95A from the Space
Telescope Science Institute, which is operated by the Association of
Universities for Research in Astronomy, Incorporated, under NASA
contract NAS5-26555.



\begin{thebibliography}{}
\bibitem[\protect\citeauthoryear{{Ajhar} et~al.}{{Ajhar}
  et~al.}{1996}]{ajhar1996}
{Ajhar}, E.~A., {Grillmair}, C.~J., {Lauer}, T.~R., {Baum}, W.~A., {Faber},
  S.~M., {Holtzman}, J.~A., {Lynds}, C.~R.,  \& {O'Neil}, J., E.~J. 1996, \aj,
  111, 1110

\bibitem[\protect\citeauthoryear{{Arp}}{{Arp}}{1959}]{arp1959}
{Arp}, H. 1959, \aj, 64, 175

\bibitem[\protect\citeauthoryear{{Arp}}{{Arp}}{1958}]{arp1958}
{Arp}, H.~C. 1958, \aj, 63, 273

\bibitem[\protect\citeauthoryear{{Barmby} et~al.}{{Barmby}
  et~al.}{2000}]{barmby2000}
{Barmby}, P., {Huchra}, J.~P., {Brodie}, J.~P., {Forbes}, D.~A., {Schroder},
  L.~L.,  \& {Grillmair}, C.~J. 2000, \aj, 119, 727

\bibitem[\protect\citeauthoryear{{Battistini} et~al.}{{Battistini}
  et~al.}{1987}]{battistini}
{Battistini}, P., {Bonoli}, F., {Braccesi}, A., {Federici}, L., {Fusi Pecci},
  F., {Marano}, B.,  \& {Borngen}, F. 1987, \aaps, 67, 447

\bibitem[\protect\citeauthoryear{{Bertelli} et~al.}{{Bertelli}
  et~al.}{1994}]{bertelli}
{Bertelli}, G., {Bressan}, A., {Chiosi}, C., {Fagotto}, F.,  \& {Nasi}, E.
  1994, \aaps, 106, 275

\bibitem[\protect\citeauthoryear{{Binney} \& {Merrifield}}{{Binney} \&
  {Merrifield}}{1998}]{textbook}
{Binney}, J.,  \& {Merrifield}, M. 1998, "Galactic astronomy" (Galactic
  astronomy / James Binney and Michael Merrifield. Princeton, NJ : Princeton
  University Press, 1998. (Princeton series in astrophysics))

\bibitem[\protect\citeauthoryear{{Bohlin} et~al.}{{Bohlin}
  et~al.}{1993}]{bohlin}
{Bohlin}, R.~C., et~al. 1993, \apj, 417, 127

\bibitem[\protect\citeauthoryear{{Brocato} et~al.}{{Brocato}
  et~al.}{1989}]{brocato1989}
{Brocato}, E., {Buonanno}, R., {Castellani}, V.,  \& {Walker}, A.~R. 1989,
  \apjs, 71, 25

\bibitem[\protect\citeauthoryear{{Dolphin}}{{Dolphin}}{1997}]{dolphin}
{Dolphin}, A. 1997, New Astronomy, 2, 397

\bibitem[\protect\citeauthoryear{{Freedman} \& {Madore}}{{Freedman} \&
  {Madore}}{1990}]{freedman}
{Freedman}, W.~L.,  \& {Madore}, B.~F. 1990, \apj, 365, 186

\bibitem[\protect\citeauthoryear{{Heasley} et~al.}{{Heasley}
  et~al.}{1988}]{heasley1988}
{Heasley}, J.~N., {Friel}, E.~D., {Christian}, C.~A.,  \& {Janes}, K.~A. 1988,
  \aj, 96, 1312

\bibitem[\protect\citeauthoryear{{Hodge}}{{Hodge}}{1961}]{hodge1961}
{Hodge}, P.~W. 1961, \apj, 133, 413

\bibitem[\protect\citeauthoryear{{Hodge} \& {Schommer}}{{Hodge} \&
  {Schommer}}{1984}]{hodge1984}
{Hodge}, P.~W.,  \& {Schommer}, R.~A. 1984, \pasp, 96, 28

\bibitem[\protect\citeauthoryear{{Holland}, {Fahlman}, \& {Richer}}{{Holland}
  et~al.}{1997}]{holland1997}
{Holland}, S., {Fahlman}, G.~G.,  \& {Richer}, H.~B. 1997, \aj, 114, 1488

\bibitem[\protect\citeauthoryear{{Holtzman} et~al.}{{Holtzman}
  et~al.}{1995}]{holtzman}
{Holtzman}, J.~A., {Burrows}, C.~J., {Casertano}, S., {Hester}, J.~J.,
  {Trauger}, J.~T., {Watson}, A.~M.,  \& {Worthey}, G. 1995, \pasp, 107, 1065

\bibitem[\protect\citeauthoryear{{Huchra}, {Kent}, \& {Brodie}}{{Huchra}
  et~al.}{1991}]{huchra1991}
{Huchra}, J.~P., {Kent}, S.~M.,  \& {Brodie}, J.~P. 1991, \apj, 370, 495

\bibitem[\protect\citeauthoryear{{Jablonka} et~al.}{{Jablonka}
  et~al.}{2000}]{jablonka2000}
{Jablonka}, P., {Courbin}, F., {Meylan}, G., {Sarajedini}, A., {Bridges},
  T.~J.,  \& {Magain}, P. 2000, \aap, 359, 131

\bibitem[\protect\citeauthoryear{Larsen \& {Richtler}}{Larsen \&
  {Richtler}}{1999}]{larsen}
Larsen, S.~S.,  \& {Richtler}, T. 1999, \aap, 345, 59

\bibitem[\protect\citeauthoryear{{Magnier} et~al.}{{Magnier}
  et~al.}{1997}]{magnier1997}
{Magnier}, E.~A., {Prins}, S., {Augusteijn}, T., {van Paradijs}, J.,  \&
  {Lewin}, W. H.~G. 1997, \aap, 326, 442

\bibitem[\protect\citeauthoryear{{Nedialkov} \& {Ivanov}}{{Nedialkov} \&
  {Ivanov}}{1999}]{nedialkov}
{Nedialkov}, P.~L.,  \& {Ivanov}, V.~D. 1999, A\&AT, 17, 367

\bibitem[\protect\citeauthoryear{{Pecci} et~al.}{{Pecci}
  et~al.}{1996}]{pecci1996}
{Pecci}, F.~F., et~al. 1996, \aj, 112, 1461

\bibitem[\protect\citeauthoryear{{Rich} et~al.}{{Rich} et~al.}{1996}]{rich1996}
{Rich}, R.~M., {Mighell}, K.~J., {Freedman}, W.~L.,  \& {Neill}, J.~D. 1996,
  \aj, 111, 768

\bibitem[\protect\citeauthoryear{{Sargent} et~al.}{{Sargent}
  et~al.}{1977}]{sargent}
{Sargent}, W. L.~W., {Kowal}, C.~T., {Hartwick}, F. D.~A.,  \& {van den Bergh},
  S. 1977, \aj, 82, 947

\bibitem[\protect\citeauthoryear{{Shapley}}{{Shapley}}{1930}]{monograph}
{Shapley}, H. 1930, Harvard Obs. Monographs No. 2 (Harvard)

\bibitem[\protect\citeauthoryear{{Stetson}, {Davis}, \& {Crabtree}}{{Stetson}
  et~al.}{1990}]{stetson}
{Stetson}, P.~B., {Davis}, L.~E.,  \& {Crabtree}, D.~R. 1990, in ASP Conf. Ser.
  8: CCDs in astronomy, 289

\bibitem[\protect\citeauthoryear{{van den Bergh}}{{van den
  Bergh}}{1969}]{vandenbergh}
{van den Bergh}, S. 1969, \apjs, 19, 145

\bibitem[\protect\citeauthoryear{{Walterbos} \& {Kennicutt}}{{Walterbos} \&
  {Kennicutt}}{1988}]{walterbos}
{Walterbos}, R.,  \& {Kennicutt}, R. 1988, \aap, p.~61


\end{thebibliography}

\clearpage

\begin{deluxetable}{ccccccccccc} 
\small
\tablecolumns{11}
\tablecaption{Global parameters of our four clusters; typical errors are given in the text.  The subscript ``c" indicates a reddening corrected magnitude.} 

\tablehead{
\colhead{\tiny{\bf{Name}}} &
\colhead{\tiny{\bf{GCD}}} &
\colhead{\tiny{\bf{$\bf{r_{1/2}}$ (pc)}}} &
\colhead{\tiny{\bf{$\bf{m_{U_{bat}}}$}}} &
\colhead{\tiny{\bf{$\bf{m_{B_{bat}}}$}}} &
\colhead{\tiny{\bf{$\bf{m_{V_{bat}}}$}}} &
\colhead{\tiny{\bf{$\bf{E_{B-V}}$}}} &
\colhead{\tiny{\bf{$\bf{m_{U_c}}$}}} &
\colhead{\tiny{\bf{$\bf{m_{B_c}}$}}} &
\colhead{\tiny{\bf{$\bf{m_{V_c}}$}}} &
\colhead{\tiny{\bf{$\bf{M_{V_c}}$}}} 
}                                  
\startdata
G38 & 14.1 & 3.7 & 16.63 & 16.50 & 16.28 & 0.31 & 14.99 & 15.36 & 15.27 & -9.16\nl
G44 & 13.2 & 3.5 & 17.81 & 17.90 & 17.71 & 0.23 & 16.34 & 16.67 & 16.71 & -7.72\nl
G94 & 10.0 & 5.3 & 18.16 & 18.17 & 17.78 & 0.20 & 16.98 & 17.41 & 17.36 & -7.07\nl
G293 & 10.4 & 2.9 & 17.93 & 18.39 & 18.18 & 0.20 & 16.95 & 17.49 & 17.45 & -6.98\nl
\hline
\enddata
\end{deluxetable}

\begin{deluxetable}{ccccc} 
\small
\tablecolumns{5}
\tablecaption{Ages and metallicities determined by isochrone fitting (columns 2 and 4) and by statical comparison of theoretically derived CMD's (columns 3 and 5).} 
\tablehead{
\colhead{\bf{Name}} &
\colhead{\bf{Age$_{\bf{iso}}$ (Myr)}} &
\colhead{\bf{\bf{Age$_{\bf{match}}$ (Myr)}}} &
\colhead{\bf{\bf{Z$_{\bf{iso}}$}}} &
\colhead{\bf{\bf{Z$_{\bf{match}}$}}}
}                                  
\startdata
G38 & $100\pm40$ & $99\pm67$ & $0.008\pm0.01$ & $0.014\pm0.02$ \nl
G44 & $100\pm40$ & $83\pm68$ & $0.008\pm0.01$ & $0.013\pm0.02$ \nl
G94 & $160\pm60$ & $136\pm54$ & $0.008\pm0.01$ &$ 0.016\pm0.02$  \nl
G293 & $63\pm15$ & $40\pm40$ & $0.02\pm0.01$ &$ 0.016\pm0.02$ \nl
\hline
\enddata
\end{deluxetable}

\begin{figure}
\centerline{\psfig{file=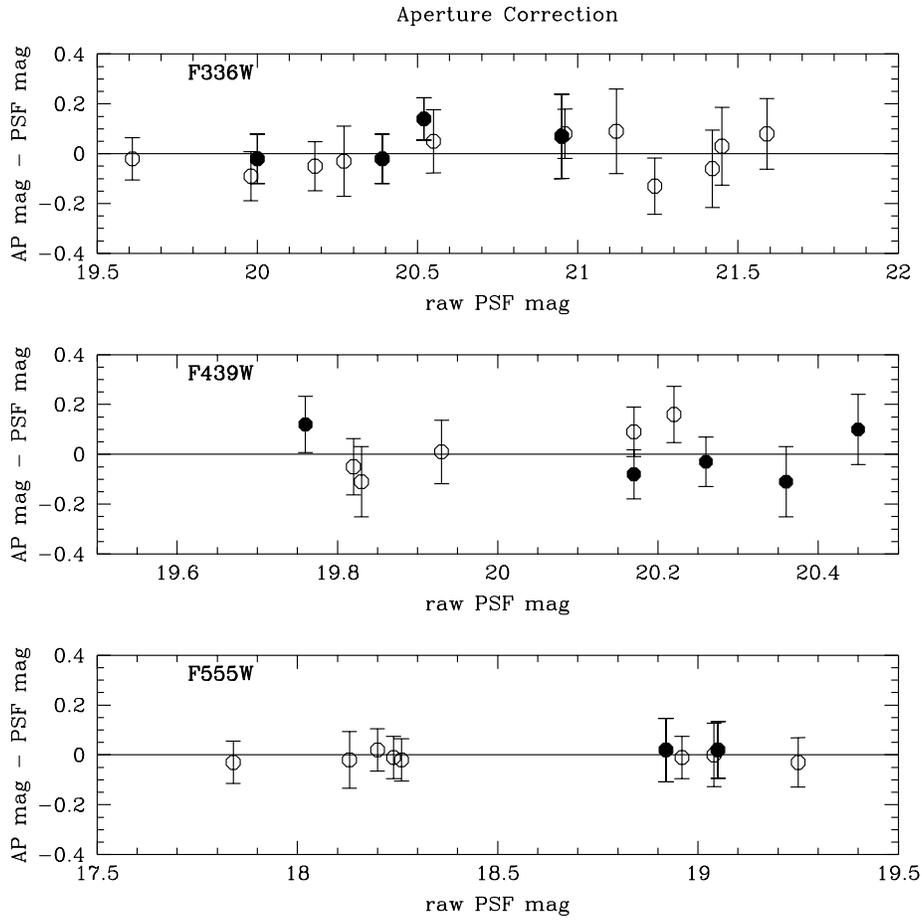,height=5.0in,angle=0}}
\caption{The residuals of the comparison of aperture magnitudes with the PSF magnitudes for the best sampled stars from the 3 bandpasses.  Filled hexagons are isolated stars. Open hexagons have subtracted neighbors within the aperture.}
\end{figure}

\begin{figure}
\centerline{\psfig{file=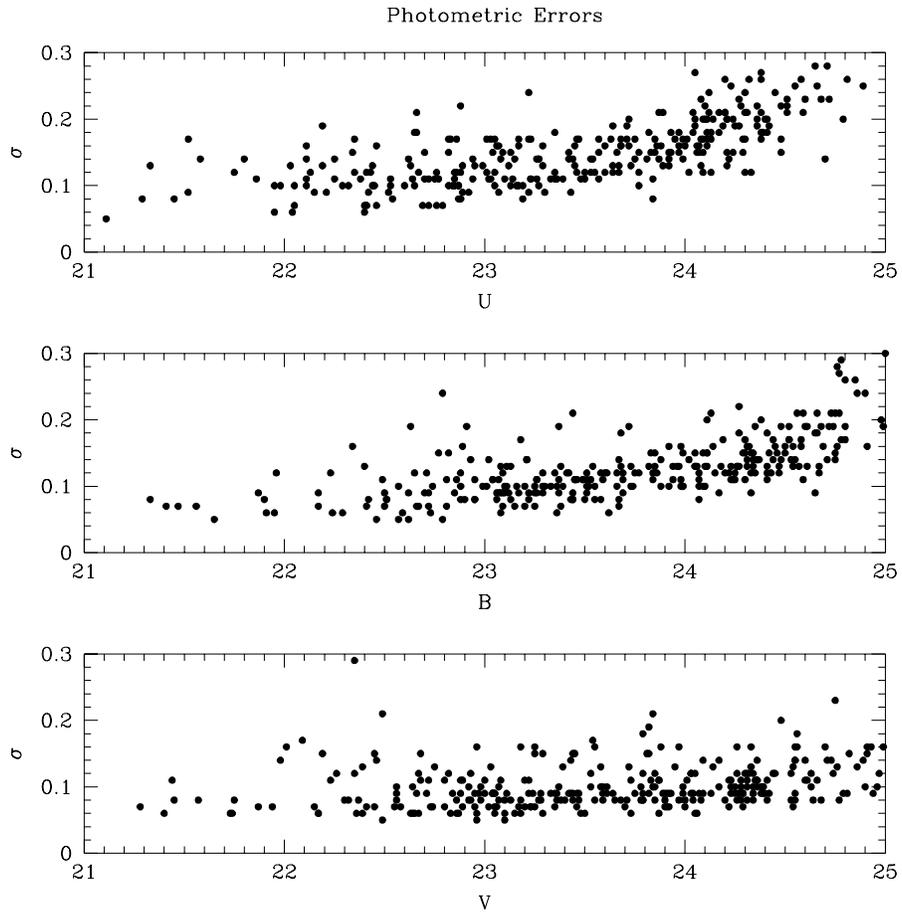,height=5.0in,angle=0}}
\caption{The calculated photometric errors for the stars in U, B, and V as a function of magnitude.}
\end{figure}

\begin{figure}
\centerline{\psfig{file=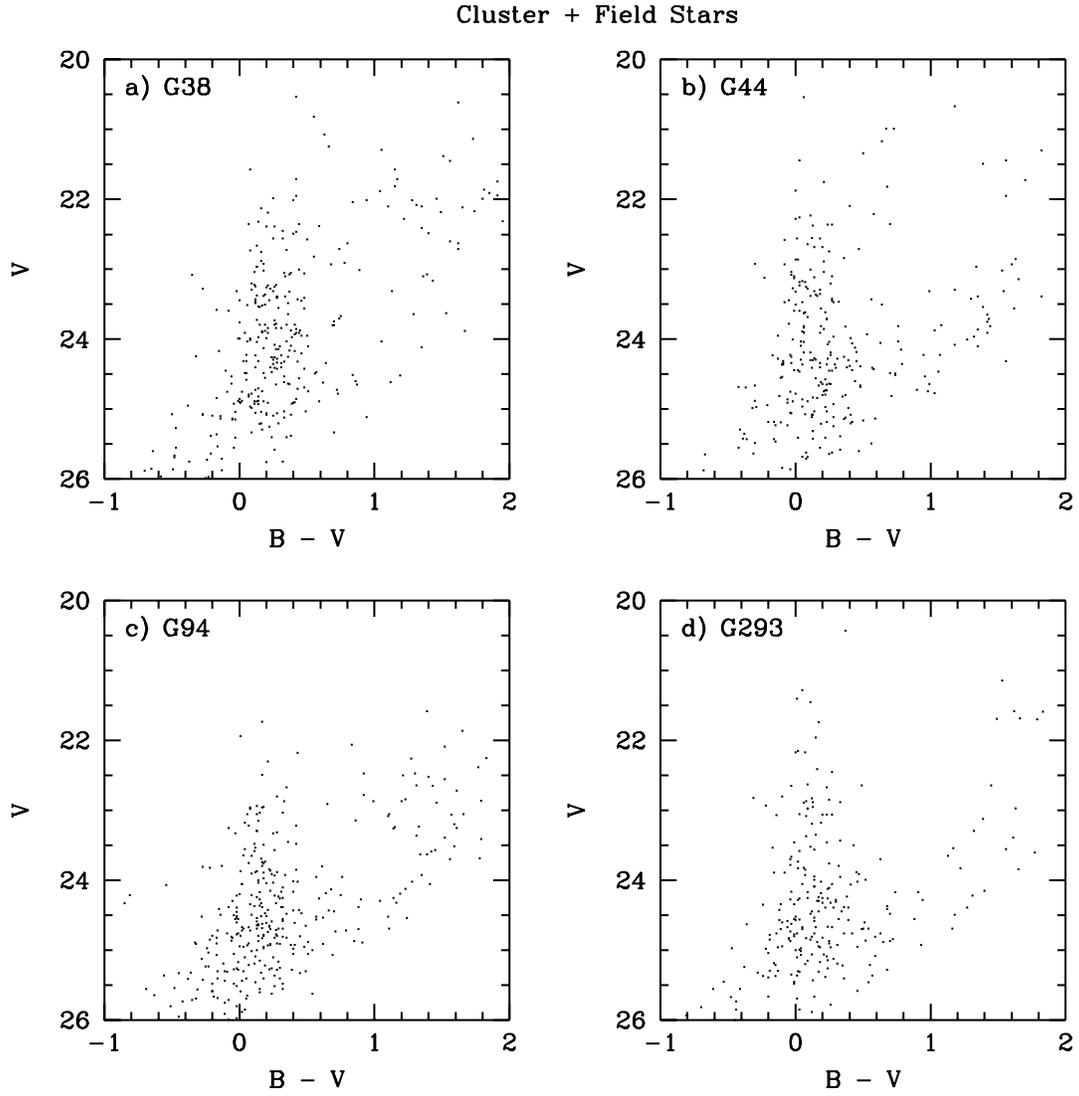,height=6.0in,angle=0}}
\caption{CMDs of the 4 clusters before field subtraction or reddening correction.  The upper main sequence is apparent near B-V = 0.  Field stars are peppered over the red side of the diagram.}
\end{figure}

\begin{figure}
\centerline{\psfig{file=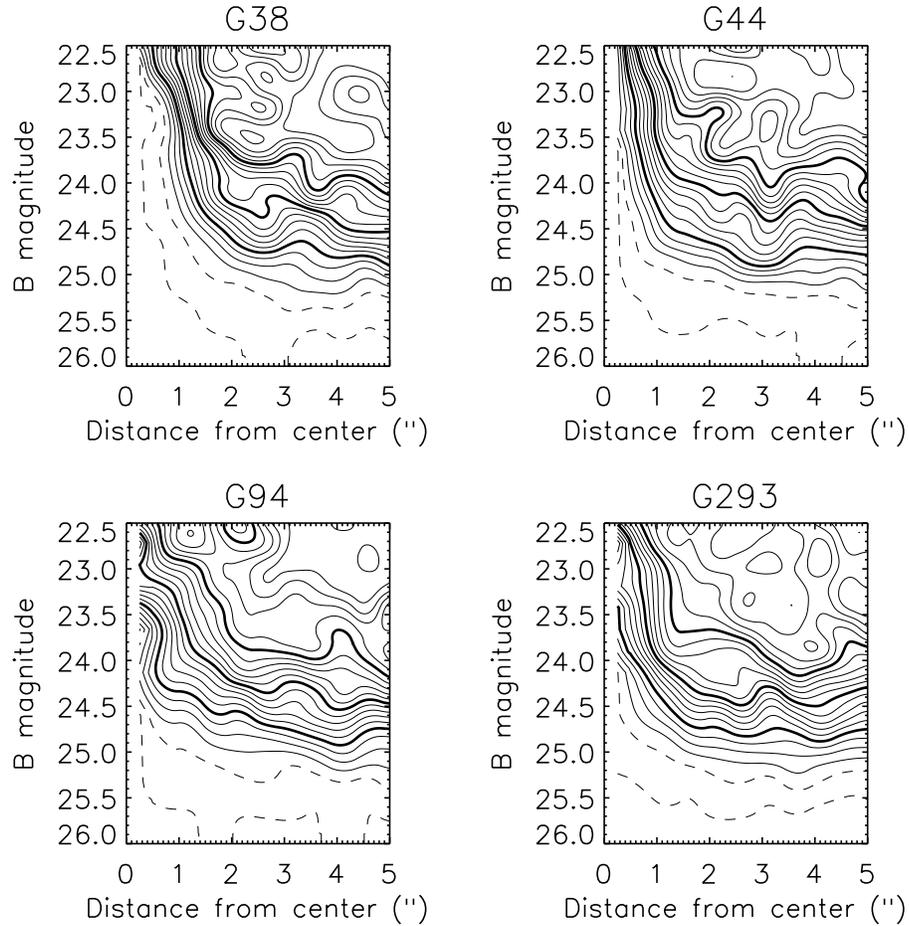,height=5.0in,angle=0}}
\caption{Completeness for the four clusters as a function of distance from the cluster center and apparent B magnitude. Dotted contours show where the completeness was less than ten percent (0.5 and 5 percent).  The first solid contour marks ten percent completeness, and successive contours each mark five percent increases in completeness, so that the thick contours mark 25, 50, and 75 percent completeness.  Above 80 percent is not as systematic due to less sampling at bright magnitudes far away from the cluster. }
\end{figure}

\begin{figure} 
\centerline{\psfig{file=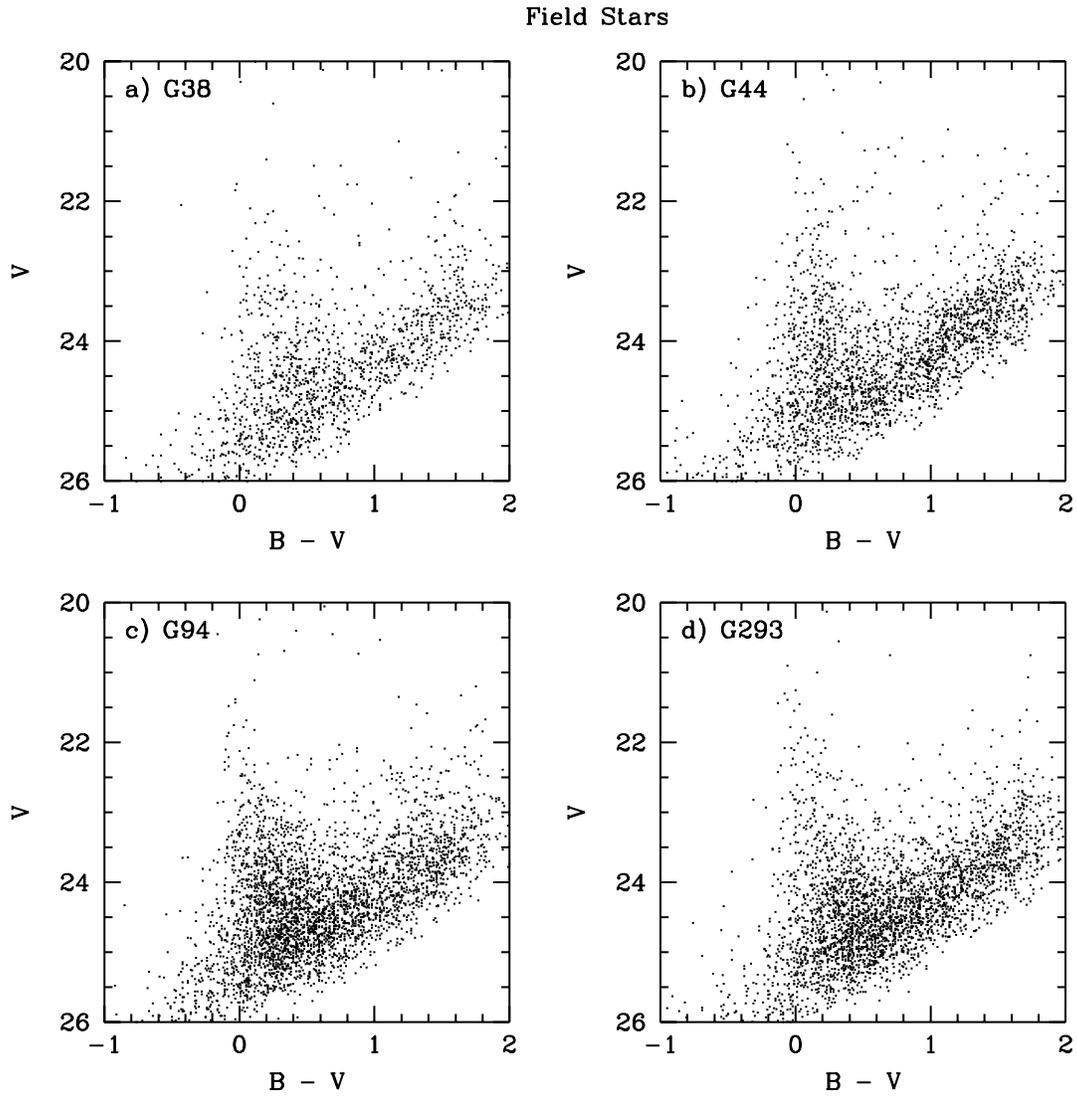,height=6.0in,angle=0}}
\caption{CMDs of stars at distances larger than 11.5" from the center
of each cluster.  Notice the smaller number of stars found around G38,
the cluster with the largest reddening value.}  
\end{figure}

\begin{figure} 
\centerline{\psfig{file=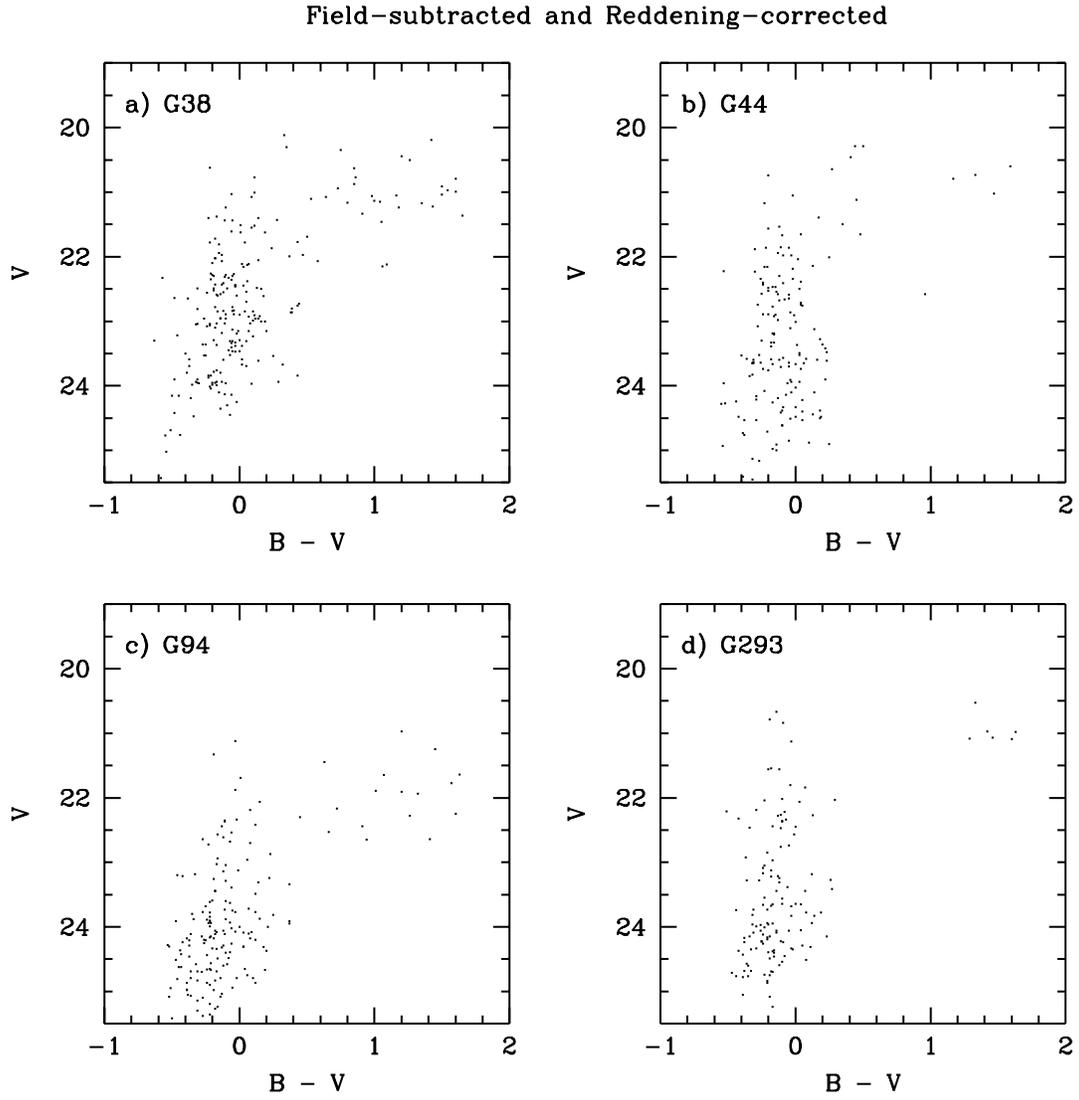,height=6.0in,angle=0}}
\caption{The field subtracted and reddening corrected cluster CMDs.
The populous main sequence and the supergiant population were not seen
in the field, and the upper main sequence was corrected to lie at the
appropriate B-V color.}  
\end{figure}

\begin{figure} 
\centerline{\psfig{file=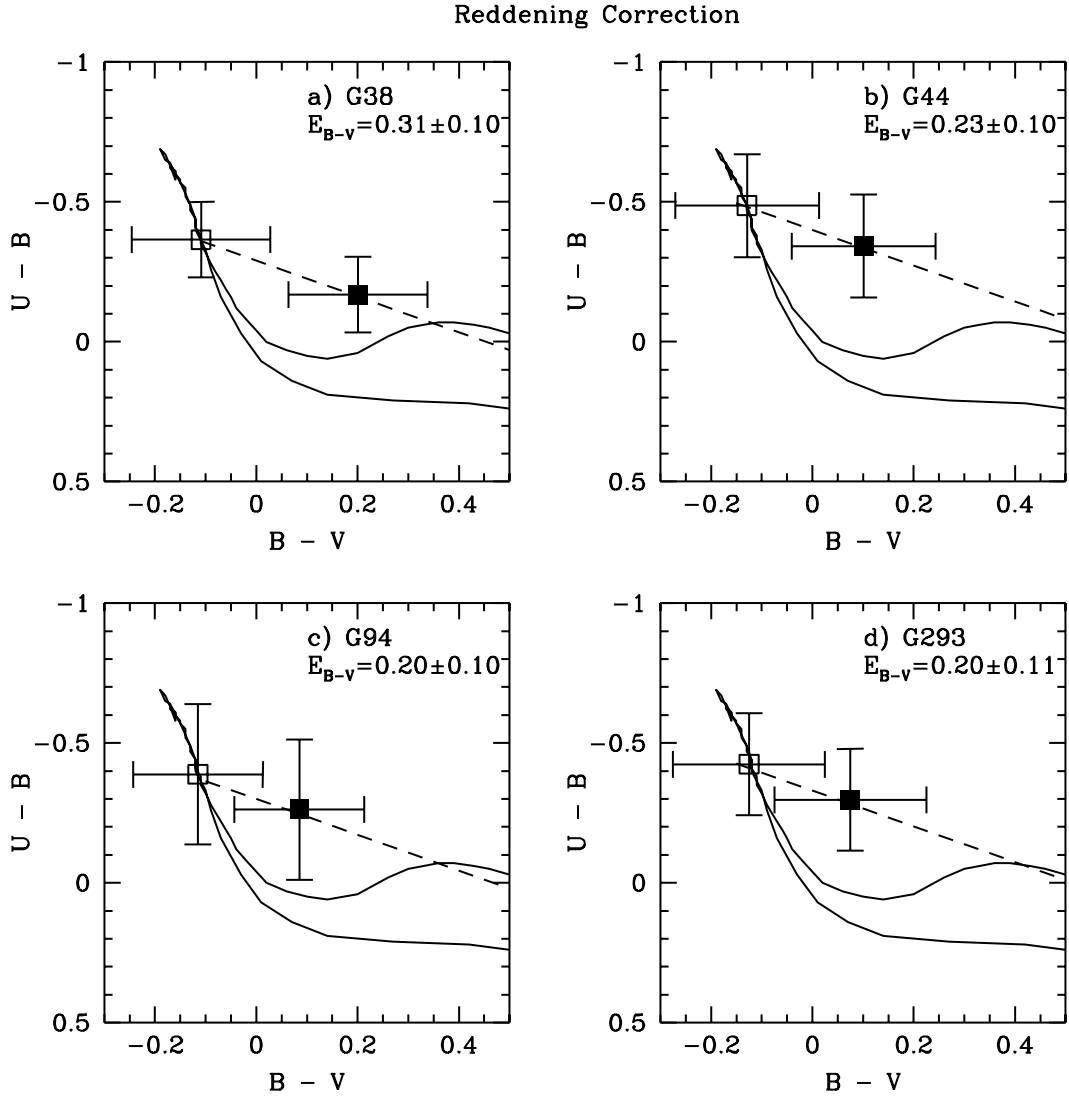,height=6.0in,angle=0}}
\caption{Reddening determination for the four clusters. Solid lines show the model colors for stars on the main sequence (top curve), and supergiants (bottom curve).  The dotted line shows the reddening line.  Closed boxes are average colors of the pre-corrected upper main sequence stars, and open boxes are the same distributions after applying our most likely reddening correction value. Our most likely reddening values and their errors are also given.}  
\end{figure}

\begin{figure} 
\centerline{\psfig{file=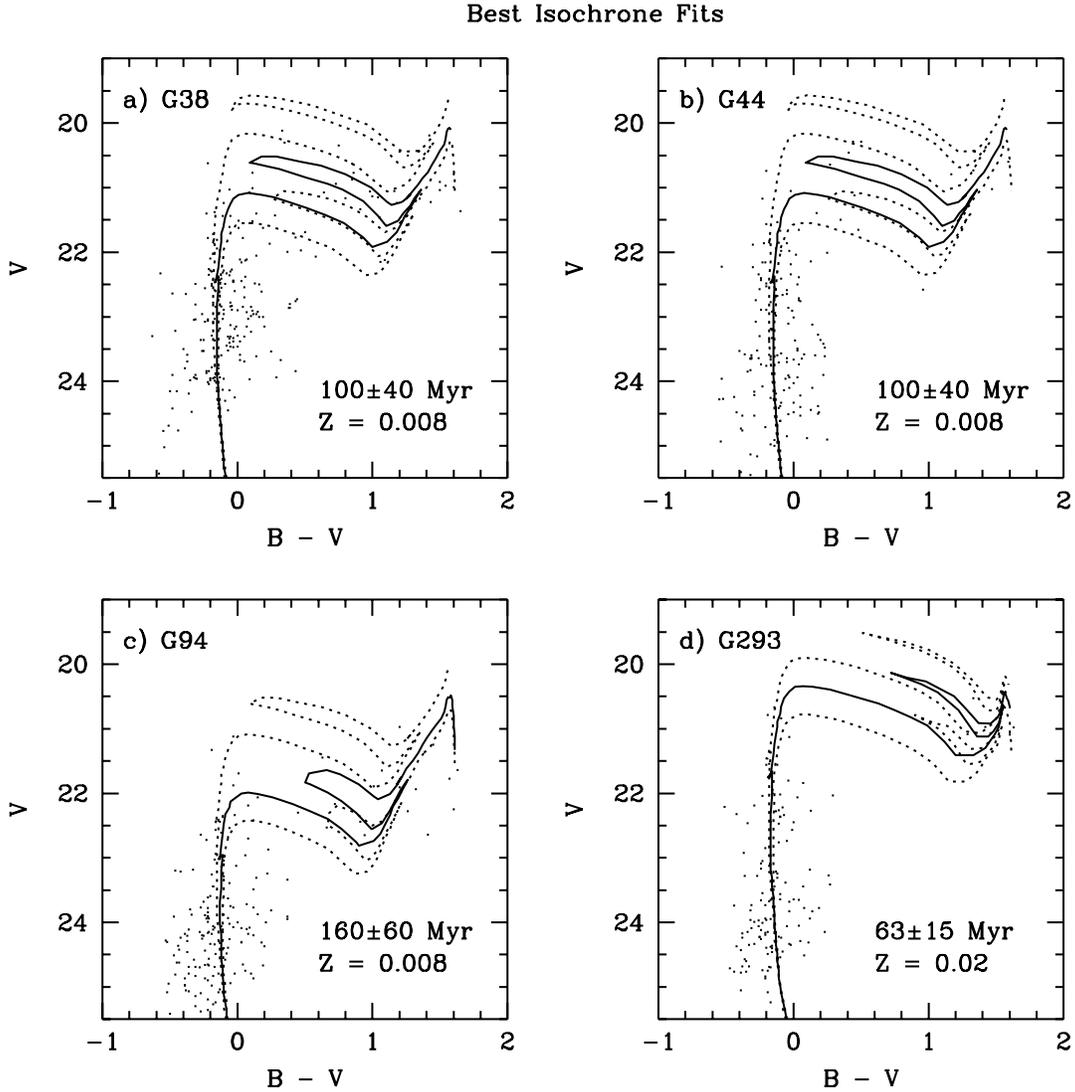,height=6.0in,angle=0}}
\caption{By eye age determination using model isochrones from Bertelli et al. (1994).  Dotted lines mark the upper and lower limits.  These were determined by finding the age where the blue loops fell below and above the location of the supergiants.  The solid isochrones mark our best fits for the turnoff age.  The metallicities of the isochrones were chosen based on the distribution of the supergiants, and may not be reliable.} 
\end{figure}

\end{document}